# Copolymer template control of gold nanoparticle synthesis via thermal annealing

*A. Plaud[1], A. Sarrazin[1], J. Béal[1], J. Proust[1], P. Royer[1], J.-L. Bijeon[1], J. Plain[1], P.-M. Adam[1] and T. Maurer[1]\**

1. Univ Technol Troyes, Lab Nanotechnol & Instrumentat Opt, Inst Charles Delaunay, CNRS UMR 6279, F-10010 Troyes, France

We present here an original process combining top-down and bottom-up approaches by annealing a thin gold film evaporated onto a hole template made by etching a PS-PMMA copolymer film. Such process allows a better control of the gold nanoparticle size distribution which provides a sharper localized surface plasmon resonance. This makes such route appealing for sensing applications since the figure of merit of the Au nanoparticles obtained after thermal evaporation is more than doubled. Such process could besides allow tuning the localized surface plasmon resonance by using copolymer with various molecular weights and thus be attractive for surface enhanced raman spectroscopy.

For the past fifteen years, the investigation of the Localized Surface Plasmon Resonance (LSPR) for plasmonic nanoparticles (NPs) has opened new perspectives for optical nanosensors(Hutter and Fendler, 2004, J.N. Anker, 2008). Indeed, LSPR sensors pave the way for size reduction compared to SPR ones (Karlsson and Stahlberg, 1995) since they give the advantage of breaking the use of prism and the need of accurate temperature control. Nevertheless, they still present difficulty in being commercialized. Therefore what is at stake today is the development of large scale arrays of plasmonic nanoparticles with tunable resonance. Top-down processes, such as Electron Beam Lithography (Haynes et al., 2003) or Focused-Ion-Etching (Tseng, 2004) found applications in the industry thanks to their high reproducibility. However it becomes expensive in time and money to fabricate structures smaller than 20nm with well-defined edges via these techniques(Pelton et al., 2008). Other synthetic routes, like colloidal synthesis(Sun and Xia, 2002, Maurer et al., 2013a) or thermal annealing after metal film evaporation(Zhu et al., 2010, Jia et al., 2012), based on the *bottom-up* approach have thus been developed. The great advantage of these approaches is that they provide significant amounts of nano-objects with dimensions smaller than 20 nm but large-scale applications are still limited due to the difficulty of organizing such metallic nano-objects over large areas(Lamarre et al., 2013).

Recently, efforts have been made to develop routes mixing top-down and bottom-up strategies like nanosphere lithography(Haynes and Van Duyne, 2001) for example. In particular, copolymer templates received great attention for their ability to drive self-assembly(Fahmi et al., 2009) but also for the possibility to form small feature size and a large variety of pattern from spheres to lamellar shapes(Krausch and Magerle, 2002, Tseng and Darling, 2010). The degradation of one of the two polymers either by Reactive Ion Etching (RIE)(Asakawa and Hiraoka, 2002, Asakawa and Fujimoto, 2005) or after UV-exposure(Kang et al., 2009, Maurer et al., 2013b) then provides arrays of nanopores. Therefore, the fabrication of copolymer templates is a bottom-up strategy which allows preparing lithography masks with long-range ordered pores. However a real difficulty remains the possibility to obtain deep enough pores(Kang et al., 2009) since their depth remains lower than few tens of nanometers.

This paper presents a process based on both bottom-up and top-down strategies to fabricate gold nanoparticles with a good control of the diameter dispersion. Among the different gold nanoparticle synthesis, metal island films have been intensively investigated for more than 20 years because of their easy way of fabrication and their ability to detect biomolecules(Vo-Dinh, 1998, Jia et al., 2012). However, this process exhibits a lack for nanoparticle size distribution control for evaporated film thickness near the percolation threshold(K. Jia et al., 2012). The idea here consists in combining advantages of copolymer templates- which provide arrays of holes with well-controlled diameters but are not deep enough for lift-off process- and of metal island films- which do not exhibit evaporation thickness issues but do not allow accurate diameter control of the nanoparticles. Therefore, this paper shows how the size distribution control of the gold islands via arrays of nanopores may be obtained after etching copolymer templates. The advantage of such materials for Localized Surface Plasmon (LSP) sensors is that they provide large areas of gold (or potentially other plasmonic) nanoparticles with narrower size distribution and thus with sharper LSP resonances and higher factor of merit for sensing. Moreover, such process paves the way for size and thus LSPR tuning of the Au NPs via control of the copolymer template domains.

Experimental: fabrication of Au nanoparticle arrays

The copolymer chosen in this study was PS-PMMA since PS and PMMA exhibit about the same glass transition temperature $T_g$ of 100-110°C. Its microphase separation has been well-extended studied and it has been shown that the order-disorder temperature $T_{ODT}$ was higher than 230°C but the chains of PMMA started decomposing at 220°C. Thus, the PS-PMMA block copolymer has to be annealed between 110°C and 210°C(Asakawa and Hiraoka, 2002, Asakawa and Fujimoto, 2005). The solvent used in this process was the Propylene Glycol Monomethyl Ether Acetate (PGMEA) (boiling point around 150°C) since it allows dissolving both PS and PMMA polymers. PS-PMMA block copolymer, with the following molecular weights $M_w(PS)$ = 96500 / $M_w(PMMA)$ = 35500 was purchased from Polymer Source. In order to get PMMA cylindrical domains into the PS matrix, the composition ratio should fit to 20% of PMMA(Asakawa and Hiraoka, 2002). PS or PMMA homopolymer was thus added depending on the copolymer characteristics to achieve such composition ratio. The polymers were dissolved in PGMEA solvent to form 3%wt solutions. The solutions were spin-coated onto glass substrates coated with an Indium Tin Oxide (ITO) layer. Then the samples are annealed under $N_2$ atmosphere during 8 hours to induce the copolymer microphase separation. The thickness of the films was assessed to 40nm. **Fig. 1A** shows the AFM phase images of the copolymer template and the corresponding size distribution of the PMMA domains which is centered on 57nm. **Fig. 1B** indicates that PMMA domains are 2-3nm higher than the PMMA matrix.

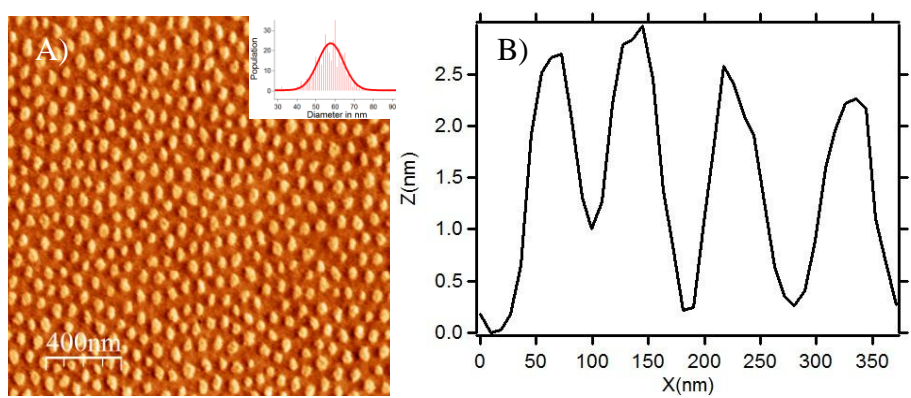

**Fig. 1** A) AFM phase images of the PS(96500)-PMMA(35500) copolymer template after microphase segregation. The PMMA domains are bright and the PS matrix is dark. The top-right inserts indicate the size distribution of the PMMA domains. B) AFM height profile indicating that the PMMA domains are 2-3nm higher than the PS matrix.

The aim was then to selectively etch PMMA domains in order to obtain arrays of nanoholes. It has already been shown that PMMA is preferably etched to PS during RIE process with $O_2$ plasma(Asakawa and Hiraoka, 2002, Asakawa and Fujimoto, 2005). The copolymer templates were first etched with a power of 150W during a short time (5s) to smooth the surface. Then, the anisotropic attack of PMMA domains is granted by an etching process with a power of 50W during a various time depending on the copolymer template thickness. Eventually, the samples are etched with a power of 10W during 30s in order to remove PMMA at the bottom of the holes. **Fig. 2A** shows how the nanoholes are formed during the RIE process described above. The as-prepared nanoholes exhibit depth of about 10nm to 35nm depending on the etching time at 50W. For this process, the hole depth was set to 20nm (see **Fig. 2B**). To get deeper holes, it is necessary to prepare thicker polymeric films. By increasing the polymer concentration in the initial polymer solution up to 6%, the thickness can be increased to 200nm. However, since the RIE is not fully anisotropic, holes start joining together during the process. Therefore other etching process than RIE should be preferred. UV-exposure of the copolymer template is a promising alternative(Maurer et al., 2013b). Moreover, the size distribution of the holes is broadened since etching may lead to the formation of larger holes when etched domains are touching if too close. Thus, the average diameter is increased to 74nm and the standard deviation is 50nm which indicates that there can be as well small holes (diameters inferior to 30nm) and very large holes as observed in **Fig. 2A**.

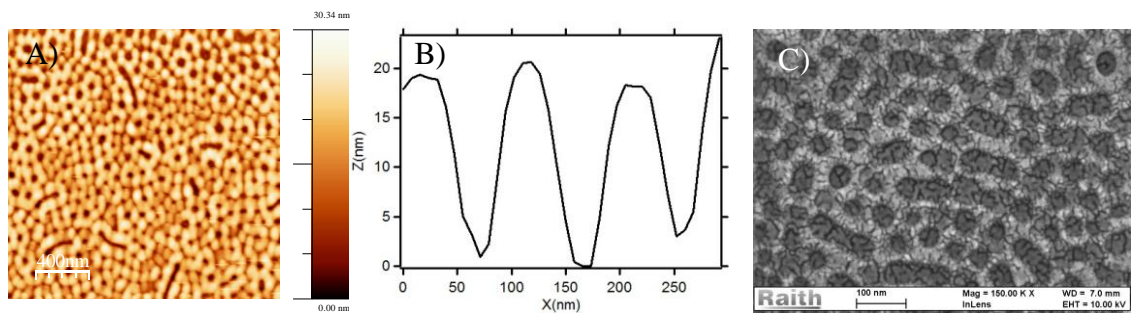

**Fig. 2** A) AFM topography images of hole arrays after Reactive Ion Etching with $O_2$ plasma. B) AFM height profile indicating a hole depth of 20nm and C) SEM image of the hole template just after Au film deposition.

In the next step, 6nm of gold film were evaporated under $10^{-5}$ Torr onto the etched copolymer templates (see Fig. 2C) but also onto glass (coated with a thin ITO layer) substrates as reference samples. Then the samples were annealed at 200°C on a hot plate for at least four hours. Finally, lift-off procedure in acetone during one night allowed removing the polymer mask and the metal on top of it.

Results and discussion

SEM images were performed for all the samples in order to illustrate the role of the lithography mask onto the size distribution of the nanoparticles (see **Fig. 3A and B**). The effect of the nanohole template on the size distribution is clearly evidenced for the samples with 6nm gold film evaporated. In this case, the annealing process without nanohole arrays does not lead to well-defined nanoparticles. It proves that the lithography masks made from the copolymer template really help the formation of gold nanoparticles. Moreover, size distributions were extracted from the SEM images and led to a size average of 43nm with a standard deviation of 12nm. The average diameter of the NPs prepared using hole templates is smaller than the average diameter of the holes probably because several NPs can be formed into the larger holes. Moreover, it proves that the use of lithography masks before gold film evaporation and annealing leads to a narrower size distribution. Note that lift-off with acetone is required to sharpen the size distribution by removing Au NPs which stand on the polymer film. To complete the characterization of as-prepared Au NPs, AFM images were made (see **Fig 3C**) and allowed assessing Au NP height to about 8nm (see **Fig. 3D**).

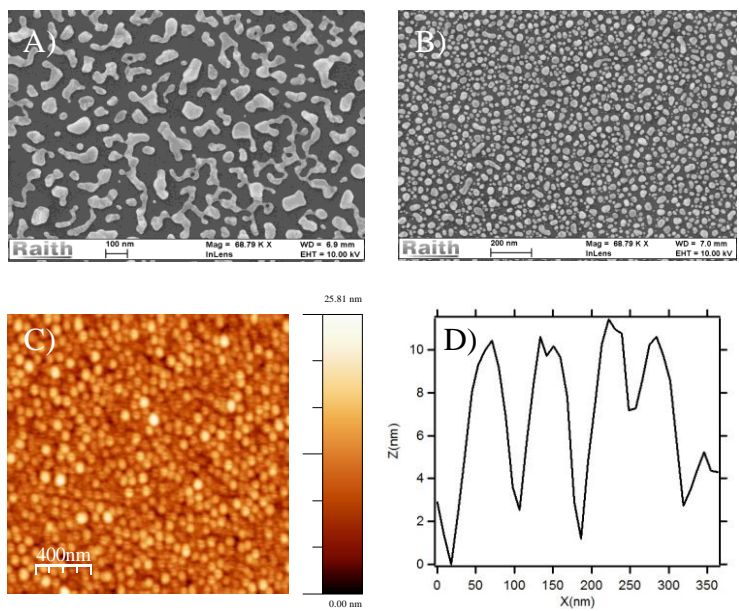

**Fig. 3** A) SEM images of Au NPs obtained after thermal annealing at 200°C during ten hours of a 6nm gold film directly deposited onto glass substrate and B) onto the hole array obtained after etching of copolymer templates, C) topography AFM image of Au NPs obtained via thermal annealing of 6nm gold film deposited onto the hole template and D) AFM height profile indicating that the as-prepared Au NPs are about 8nm high.

The optical properties of the Au NPs obtained after thermal annealing were studied via optical extinction measurement with a transmission optical microscope coupled to a micro-spectrometer by a multimode optical fiber. **Fig. 4** provides the extinction spectra for Au NPs obtained after thermal annealing using the hole template or not and whose SEM images are provided in Figure 4. The LSPR resonance of Au films obtained after thermal annealing without using the hole template is centered on 592nm and exhibit a width of 144nm. The use of hole templates in the annealing process leads to a sharper better LSPR peak since it is located at 569nm with a width of 87nm. Note that the LSPR peak is blue-shifted when using the hole template due to the smaller size of the as-prepared Au NPs(Hutter and Fendler, 2004, Pelton et al., 2008).

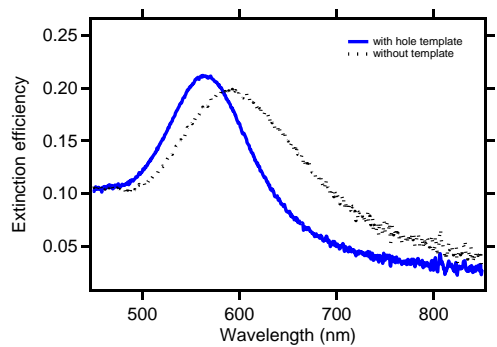

**Fig. 4** Optical extinction spectra of Au NPs using hole template (full line) or not (dashed line). The use of hole templates induces sharpening of the LSPR peak.

Sharpening LSPR peaks reveals to be appealing for sensing applications since potential for LSP sensors lies in their factor of merit (FoM) which is given by the ratio between the sensitivity of the LSP mode- that is to say how much it is shifted per refractive index unit (RIU) of the surrounding medium- and its full width at half maximu (FWHM). To probe such potential, sensing measurements were performed by depositing mixtures of water (refractive index of 1.33) and glycerol (refractive index of 1.47) which gives access to the sensitivity of the LSP mode. Figure 5 displays the extinction spectra when water or glycerol is deposited. For Au NPs made without using the hole template, the shift of the LSP mode is 11nm which leads to a value of sensitivity equal to 78nm/RIU and thus a FoM of 0.54. When the Au NP were prepared with the hole template, the sensitivity is increased to 15nm which corresponds to a sensitivity of 107nm either a FoM of 1.32. Therefore, the use of the hole template allows increasing the FoM of the LSP sensor more than twice higher.

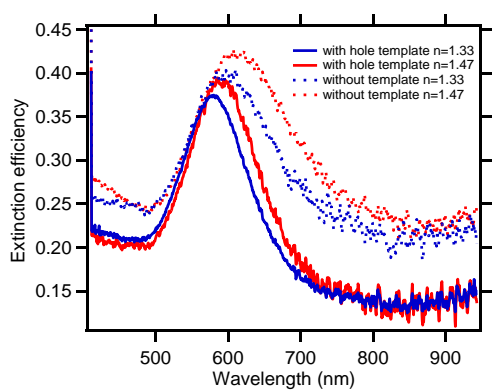

**Fig. 5** Optical extinction spectra of Au NPs using hole template (full line) or not (dashed line) when water (blue) or glycerol (red) is deposited. Spectra for mixtures with different ratio are not shown to keep the figure clear.

The process described in this paper shows potential for improving performances of LSP sensors made by thermal annealing of gold film. Such devices are usually attractive for their easy way of making but suffer from a lack of sensitivity and broad resonances when the evaporated metal film thickness is above 2nm(K. Jia et al., 2012). In order to bring LSPR substrates to the market, efforts have been made to develop synthetic routes combining large scale fabrication, low cost and tunable optical properties. It is here shown that using hole templates made from copolymer mask etching conducts to a better control of the Au NPs size distribution even for thicker evaporated metal film. The main advantages of this original process are (i) to provide sharper LSP resonances and (ii) a low annealing temperature compared to similar routes (K. Jia et al., 2012). In order to improve this process, efforts will be focused onto the tunability of the LSPR. Two ways should be considered: i) adjusting the copolymer molecular weight and ii) etching the copolymer template with UV-exposure which is a more anisotropic process and thus allows using thicker initial copolymer masks. In case of success, this could be attractive in the next future to design Surface Enhanced Raman Scattering (SERS) sensors tunable to the excitation laser wavelength.

**Acknowledgments**

Financial support of NanoMat (www.nanomat.eu) by the "Ministère de l'enseignement supérieur et de la recherche," the "Conseil régional Champagne-Ardenne," the "Fonds Européen de Développement Régional (FEDER) fund," and the "Conseil général de l'Aube" is acknowledged. T. M thanks the DRRT (Délégation Régionale à la Recherche et à la Technologie) of Champagne-Ardenne, the EPF and the UTT via the strategic program « CODEN », the Labex ACTION project (contract ANR-11-LABX-01-01) and the CNRS via the chaire « optical nanosensors » for financial support.